\documentclass{iaus}
\usepackage{graphicx}

\title[Random Magnetic Fields in the ISM] %% give here short title %%
{The Role of the Random Magnetic Fields in the ISM: HVCs Numerical Simulations}

\author[Santill\'an \etal]   %% give here short author list %%
{A. Santill\'an$^1$%
  \thanks{Present address: Unidad de Investigaci\'on en C\'omputo 
Aplicado--DGSCA, UNAM, 04510, Mexico City, Mexico},
J. Kim$^2$, F.J. S\'anchez--Salcedo$^3$ \break 
J. Franco$^3$ \and L. Hern\'andez--Cervantes$^3$}

\affiliation{$^1$Direcci\'on General de Servicios de C\'omputo Acad\'emico, 
UNAM, 04510, Mexico City, Mexico \break email: alfredo@astrosu.unam.mx\\[\affilskip]
$^2$Korea Astronomy and Space Science Institute, 61--1, Hwaam--dong, 
Yuseong--gu, Daejeon, Republic of Korea 305-348\\[\affilskip]
$^3$Instituto de Astronom\'\i a , UNAM, 04510, Mexico City, Mexico}

\pubyear{2009}
\volume{259}  %% insert here IAU Symposium No.
\pagerange{119--126}
\date{?? and in revised form ??}
\jname{Cosmic Magnetic Fields: From Planets, to Stars and Galaxies}
\editors{K.G. Strassmeier, A.G. Kosovichev \& J. Beckmann, eds.}
\begin{document}

\maketitle

\begin{abstract}

We know that the galactic magnetic field possesses a random component in 
addition to the mean uniform component, with comparable strength of the two 
components. This random component is considered to play important roles in the 
evolution of the interstellar medium (ISM). In this work we present numerical 
simulations associated with the interaction of the supersonic flows located at 
high latitude in our Galaxy (High Velocity Clouds, HVC) with the magnetized 
galactic ISM in order to study the effect that produces a random magnetic field 
in the evolution of this objects. 

%% add here a maximum of 10 keywords, to be taken form the file <Keywords.txt>

\end{abstract}

\firstsection % if your document starts with a section,
              % remove some space above using this command.
\section{Introduction}

Numerical simulations of the evolution of HVC collisions with the Milky Way
have been performed for more than two decades by different authors. The 
details of resulting supersonic flows depend of on the model assumptions, and 
the intensity and initial configuration of the magnetic field. 
Santill\'an et al. (1999) made models that illustrate the effects of magnetic 
pressure, and differentiate them from those due to magnetic tension. 
The evolution of the interaction of a HVC with a magnetized interstellar medium,
is studied by setting a random magnetic field that satisfies the 
divergence--free constraint ($\nabla~\bullet$ $\bf B$ =0) at all times. 
To mimic the average galactic magnetic field, that is oriented parallel to
the disk, the horizontal component dominates over the vertical component. 

\section{Results}

\begin{figure}
\includegraphics[height=3in,width=5.3in]{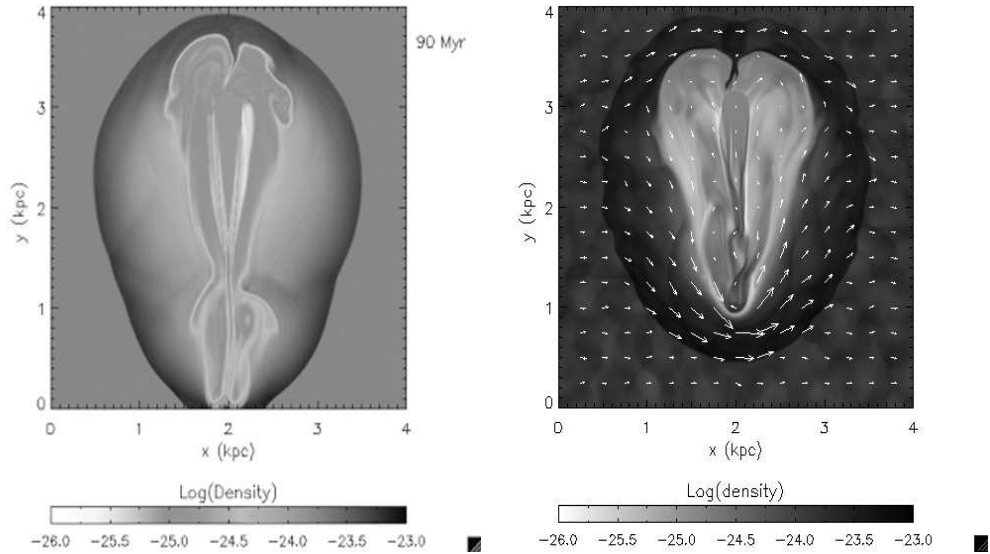}
  \caption{
Evolution of HVC in two medium: without and with magnetic field. The figure 
shows the density (\textit{gray logarithmic scale}) and the magnetic field 
indicated by \textit{arrows} at 90 Myr.}\label{fig:hvc}
\end{figure}

We perform simulations of HVCs interacting with the interstellar medium without 
and with magnetic field, ISM--1 and ISM--2, respectively, using the MHD code 
ZEUS--3D \cite{StoneNorman1992a}. The ISM--models are plane parallel and have
constant density and temperature, $n = 1~{\rm cm^{-3}}$ and $T = 1000$ K. The 
galactic gas is initially at rest.  For ISM--2 case the total intensity of
the magnetic field is 2~$\mu$G. 
The enter position of the HVC is located 3 kpc above from the midplane and has 
a velocity of 100 km/s. In the figure~\ref{fig:hvc}, the density is shown in 
logarithmic gray--scale plots and the magnetic field is indicated by arrows. 

The early times evolution for both models display the same characteristics, 
\eg the interaction between the HVC and the halo gas creates a strong galactic 
shock directed downwards, along with reverse shock that penetrates into the 
cloud. 
The galactic shock tends to move radially away from the location of impact 
\cite{Santillan1999}. As time proceeds, the structure of the HVC changed 
slightly; in both models a large fraction of the original cloud mass remains 
locked up in the shocked layer, and a small amount of it re--expanded back 
into the rear wake and tail. However, in the HD--case, the clouds moves a
greater distance than the MHD--case, due to the effect of the magnetic tension 
of the horizontal--component of random magnetic field. In the magnetic case, 
the HVC distorts and compresses the B--field lines during the evolution, 
increasing 
both the field pressure and the tension, and forming a magnetic barrier for 
the moving gas. Finally, as seen in figure~\ref{fig:hvc}, the late times 
structure produced by hydrodynamic simulations is completely different from 
that of its magnetic counterpart. In the case non--magnetic case, ISM--1, the interaction of HVC with ambient medium creates a thin structure and the size of 
perturbation region at 90 Myr has grown to nearly 3 kpc. On the other hand, for the magnetic case, the evolution of the cloud at the same evolutionary time
creates a thick structure of $\sim$2 kpc and a great amount of gas collects in 
a magnetic valley formed by the interaction with the horizontal--component of 
the random magnetic field. 

The numerical calculations were performed using UNAM's supercomputers.

\begin{acknowledgments}
This work has been partially supported from DGAPA-UNAM grant IN104306 and 
CONACyT proyect CB2006--60526.
\end{acknowledgments}

\end{document}